\documentclass[pra,
a4paper,
showpacs,
twocolumn,
superscriptaddress]{revtex4-1}

\usepackage{amssymb}
\usepackage{amsmath}
\usepackage{amsfonts}
\usepackage{braket}
\usepackage{graphicx}
\usepackage{bm}
\usepackage{color}
\usepackage{multirow}
\usepackage{natbib}
\usepackage{hyperref}
\usepackage[normalem]{ulem}
\usepackage{verbatim}
\usepackage{dcolumn}
\usepackage{ulem}
\usepackage{float}

\usepackage[dvipsnames]{xcolor}

\def \be {\begin{equation}}
\def \ee {\end{equation}}
\def \bea {\begin{align}}
\def \eea {\end{align}}

\def \BEA {\begin{eqnarray}}
\def \EEA {\end{eqnarray}}

\def \BC {\begin{cases}}
\def \EC {\end{cases}}

\begin{document}
\title
{ 
 Lateral plasmonic superlattice in  strongly dissipative regime 
}

\author{I.\,V.~Gorbenko}
\address{Ioffe Institute,
194021 St.~Petersburg, Russia}
\author{V.\,Yu.~Kachorovskii }
\address{Ioffe Institute,
194021 St.~Petersburg, Russia}


\begin{abstract} 
We calculate  transmission coefficient, $\mathcal T,$  of terahertz radiation through  lateral plasmonic superlattice  with a unit cell consisting of two regions with different plasma wave velocities, $s_1$ and $s_2$ ($s_1 > s_2$). We generalize theory developed earlier for resonant case to  the non-resonant regime, when the scattering rate, $\gamma,$  is large compared to fundamental gate-tunable frequencies $\omega_{1,2}$ of plasma oscillations in both regions. We find that absorption, and consequently $\mathcal T$, strongly depends on density modulation amplitude and on the frequency of the incoming radiation.  We describe  evolution of the absorption with increasing of radiation frequency from   the  quasi-static  regime of very low frequency to the high-frequency regime,      
     identify several dissipation regimes and find analytical expression for absorption, and, accordingly, for $\mathcal T,$ in      these regimes. A  general phase  diagram of non-resonant regime in the plane $(\omega,\omega_2)$ for fixed  $\omega_1$ is constructed.  Most importantly,  $\mathcal T$  sharply depends on the gate voltages and frequency.  In particular, for $\omega_2  \ll \omega_1,$  $\mathcal T$ strongly varies on the very small frequency scale,  $\delta \omega \ll \gamma,$ determined by the Maxwell relaxation, $\delta \omega \sim \omega_1^2/\gamma,$ so that the superlattice shows high responsivity  within  the frequency band $\delta \omega.$       

\end{abstract}

\maketitle

\section{Introduction}

Exploration of plasma waves in two-dimensional (2D) systems, which began in 1970s
\cite{chap1972,allen1977,Theis1977a, Tsui1978, Theis1978, Theis1980, Tsui1980a, Kotthaus1988}, was  initiated by theoretical analysis of spectrum of 2D plasma waves  \cite{chap1972}  and  first experimental observation of 2D plasmons in infrared light absorption in 2D electron gas with grating couplers \cite{allen1977}.   

 Plasmonic oscillations are not only of fundamental interest, but  are also very  promising for applications (for review, see  \cite{Maier2007}).  Remarkably, in 2D gated structures,   the  plasma wave velocity $s$ can be tuned  by the gate voltage  and its typical value, $s \sim 10^8$ cm/s,  exceeds  by the order  of magnitude the electron drift  velocity $\lesssim 10^7$ cm/s.  
Hence, for semiconductor structures with  submicron sizes, the frequency of plasmon oscillations falls in the terahertz (THz) range. This opens a wide avenue to create tunable semiconductor devices operating in the THz range of frequencies. 

A significant boost supporting  the idea of   2D   THz plasmonics  was given by Dyakonov and Shur, who  proposed  dc-current-driven plasma waves instability mechanism in the field-effect transistor (FET)  \cite{Dyakonov1993} and have also  shown that  FET can be used as  tunable detector, resonant or non-resonant,   of the THz radiation \cite{Dyakonov1996}. These  works initiated a series of publications focused  on the theoretical, experimental and numerical  study of plasma wave phenomena in FETs  (see reviews  \cite{Otsuji2014, Akter2021,Shur2021} and references  therein).         

Quite soon  it became clear that multi-gated 2D structures, where electron density is modulated with period $L$ (typical value $\sim 1~\mu \rm{m} $),   have significant advantages as compared  to single FETs.  In particular, for  a given value of  $L,$ coupling of system with THz radiation (with a typical wavelength $100 ~\mu \rm{m} $) increases with number of gratings, i.e. number of modulation periods. Experimentally, in these structures transmission coefficient shows plasmonic resonances of good quality up to 170 K in AlGaN/GaN-based structures \cite{Muravjov2010} and up to the room temperature in grating-gate graphene structures \cite{Boubanga-Tombet2020}. 

The grating-gate FETs are of particular interest also from the fundamental point of view. The periodical density modulation creates crystal structure for plasma waves with pass and stop bands tunable by the  gate voltage.  Hence,  such structure represents  lateral  plasmonic crystal (LPC) with the unique possibility of   pure electrical   control of  the band structure. Such LPC are actively studied theoretically and  numerically   \cite{Kachorovskii2012,Petrov2017,Aizin2023, Popov2011, Popov2015, Fateev2010, Fateev2019a},  as well as experimentally \cite{Otsuji2008,Muravjov2010, Otsuji2013EmissionGraphene,Boubanga-Tombet2020,Sai2023}.  Very recently, the possibility of tuning LPC  from weak to strong coupling regime by gate voltage  was verified  experimentally  \cite{Sai2023} and described  theoretically \cite{Gorbenko2024},  with a    good agreement between experiment and  theory.

The  study  of non-resonant regime of LPC operation, when widths of pass- and stop-bands of LPC are smaller than  the momentum relaxation rate, is  also of  very significant interest. It is worth noting that the most intense response in single FETs is usually  experimentally obtained in the non-resonant regime, when FET channel is depleted, as was first demonstrated in Ref.~\cite{Knap2002} (for review of more recent publications discussing nonresonant regime,  see Ref.~\cite{Zak2014, Vicarelli2012, Bandurin2018b, DelgadoNotario2020, DelgadoNotario2022, Soltani2020}).   Moreover, in the non-resonant regime,  a  giant dc-current-induced increase of direct voltage response \cite{Veksler2006} 
and polarization-sensitive plasma wave interference effects  \cite{Drexler2012,Romanov2013,Gorbenko2019, Gorbenko2019b}   
were  demonstrated  both theoretically and experimentally.  Also, the  appearance of dc current in modulated systems without center of inversion---the so called ratchet effect---is actively studied in LPC systems, in particular, in  the non-resonant regime \cite{Ivchenko2011,Popov2011,Rozhansky2015,Faltermeier2017,Faltermeier2018,Hubmann2020,Sai2021,Monch2022,Monch2023}.  As for  the transmission coefficient through such a crystal,  even for good structures, where  it  shows good plasmon resonances for lowest plasmonic modes, the   high-order resonances are significantly suppressed by dissipation  \cite{Muravjov2010,Boubanga-Tombet2020, Sai2023},  so that they should be described by non-resonant theory.

In this paper we calculate  transmission coefficient connected with  absorption of the THz radiation in lateral plasmonic crystal operating in the non-resonant regime. We identify several dissipation regimes and find analytical expression for absorption, and, consequently, for transmission coefficient in      these regimes.  

\section{Model}
\label{Sec-approach}
\subsection{Problem formulation    and general approach}

We follow recent works \cite{Kachorovskii2012,Petrov2017, Boubanga-Tombet2020, Gorbenko2024} and describe lateral plasmonic crystal as a system of alternating stripes
with a unit cell consisting from two regions with lengths $L_1$ and $L_2$  ($L_1+L_2=L$)  having plasma wave velocities $s_1$ and $s_2$ ($s_1>s_2$), and     operating in the  non-resonant regime: $\gamma \gg \omega_1, ~ \gamma \gg \omega_2,$  where $\gamma$  is the momentum relaxation rate and  $\omega_{1,2} = \pi s_{1,2}/L_{1,2}$.  For such  conditions,  plasma resonance are damped. 
 The purpose of this work is to  construct a  general phase  diagram of non-resonant regime in the plane $(\omega,\omega_2)$ (for fixed  $\omega_1$)  depending on gate-tunable frequencies $(s_1/ L_1,s_2/L_2). $ 
 To be specific, we fix $s_1$ and assume that    $s_2$   can be tuned by gate in the interval $0<s_2<s_1$  (hence $\omega_2$ changes from $0$ to $\omega_1$ if we put $L_1 = L_2$). Following Ref.~\cite{Boubanga-Tombet2020},  we refer to region ``1''   as {\it active} and to region ``2'' – as {\it passive}. The case $|s_1 -s_2| \ll s_1$ corresponds to the weak coupling of the LPC, when electron  density is weakly modulated.  The opposite case of the strong coupling, i.e. strong density modulation, is realized for $s_2 \ll s_1.$

We assume that system is illuminated with THz radiation, which can be assumed homogeneous due to large wavelength, and calculate transmission spectrum up to the first order of $\sigma/c$ with $c$ – the speed of light and $\sigma$ – 2D electron liquid conductivity.

In this case, one can neglect radiative decay of plasmonic oscillations (see Refs.~\cite{Mikhailov1998, Boubanga-Tombet2020}) and 
     transmission coefficient is close  to unity with a small correction, which  
        is  fully expressed in terms of Ohmic dissipation in the channel, $P$:  
\be \mathcal T  \approx 1-\frac{8 \pi \, P}{c \sqrt{\epsilon} E_0^2 },
\label{T}
\ee 
where $E_0$ is the amplitude of the incoming radiation,   and 
  $\epsilon$ is the dielectric constant.

Local dissipation within hydrodynamic approximation is given by $ m |\mathbf v (x,t) |^2/\tau$ \cite{Rozhansky2015},  where 
$\mathbf v (x,t)$ is the hydrodynamic velocity and $\tau$ is the momentum relaxation  time.
Hence, the         
ohmic dissipation per unit area in a plasmonic crystal reads 
\begin{equation}
    P = \,  \left< N \frac{m |\mathbf v (x,t) |^2}{\tau} \right>_{x,t}
\label{Eq-diss0}
\end{equation}
where $N=N(x,t)$ is the electron concentration in the channel and  $\left<\dots\right>_{x,t}$ stands for  averaging taken over area of the crystal and period of the incoming radiation.

Now let us discuss external field properties. Grating gate modulates not only electron concentration but also amplitude of electromagnetic field. So the field in the channel is a sum of homogeneous plane wave and evanescent waves modulated with a period of grating $L = L_1 + L_2$.

Effect of grating field modulation,  was discussed in details for resonant and  so-called super-resonant regimes \cite{Gorbenko2024} within the simplified model proposed in  Ref.\cite{Ivchenko2011}.  
Most importantly, due to the inhomogeneous component, a number of    {\it dark}  plasmonic modes, which are not coupled to homogeneous field, can be excited.

The dark modes have not yet been observed experimentally neither in FETs, nor in LPC. Moreover we have  recently shown theoretically  \cite{Gorbenko2024} that such modes can be only excited provided that field modulation violates inversion symmetry of the whole system (this effect can be taken into account by  introducing a  phase shift between modulations of the external field and  concentration).  For the case when  phase shift is zero,   dark modes are not excited and  field modulation  only leads  to   small corrections to amplitudes of bright modes.  Therefore, as first approximation, one can neglect these corrections and consider homogeneous excitation  $E(t) = E_0 \cos{(\omega t)}$.

\subsection{Hydrodynamic approximation} \label{HD}

In this paper, we describe LPC by using hydrodynamic approximation, so that we assume  that the  electron-electron collisions    dominate over impurity and phonon scattering and  2D electron liquid in the  channel  can  be described  by   Euler  and continuity equations.
\begin{align}
&\frac{\partial v}{\partial t}\! + \!v \frac{\partial v}{\partial x}\! \!+\! \gamma v\!=\! - \frac{e}{m} \frac{\partial U}{\partial x} + \frac{F}{m},
\label{Eq-Navier_Stokes}
\\ 
&\frac{\partial U}{\partial t} + \frac{\partial}{\partial x} (U v) = 0,
\label{Eq-continuity}
\end{align}
with local drift velocity $v(x,t,)$, and local  gate-to-channel voltage $U(x,t).$  Equilibrium   electron concentration  in the channel is controlled with gate voltage $U_g$:
\begin{equation}
 N_0=\frac{C (U_{\rm g}- U_{\rm th})}{e},    
 \label{Eq-N0}
\end{equation}
while the local concentration is given by the same equation with replacement   $U_{\rm g } \to U(x,t).$  Here $U_{\rm th}$ is the threshold voltage,  $C=\epsilon/4\pi d $ is the channel capacitance per unit area,  and $d$ is the spacer width. 
Plasmonic effects are  connected with the term ${(e/m
)} {\partial U}/{\partial x}$ in Eq.~\eqref{Eq-Navier_Stokes}. 
The  plasma wave velocity depends on the concentration  and, consequently,  is also gate-tunable:
\begin{equation} 
s = \sqrt{\frac{e (U_g-U_{th})}{m}}. 
\label{Eq-ss}
\end{equation}
%
Standard   boundary conditions between active and passive regions \cite{Kachorovskii2012, Petrov2017}  correspond to conservation of  the current, $N v,$ and energy flux, $mv^2/2+ eU.$ 
We neglect in  hydrodynamic  equations viscosity, $\eta,$ assuming that  $\eta q^2 \ll \gamma  $ (here $q \sim 1/L$), so that impurity scattering dominates over viscous friction.  
The calculation are rather standard, so we present here the results, while the specific details can be found in Ref.\cite{Gorbenko2024}.
Direct calculation of dissipation yields 
\be
\begin{aligned}
&P = \frac{ \gamma^2}{\omega^2+\gamma^2 } \left[ P_{0} + P_{\rm res} \right],
\end{aligned}
\label{Eq-maindis}
\ee
where
\be
\begin{aligned}
&P_0 = \frac{ F_0^2 C}{2 e^2 \gamma (L_1+L_2)} \left( L_1 s_1^2+L_2 s_2^2 \right),
\end{aligned}
\label{Eq-P0}
\ee
\be
\begin{aligned}
&P_{\rm res} = \frac{ F_0^2 C}{2 e^2 (L_1+L_2)} \frac{(s_1^2 - s_2^2)^2  \rm{Re}\left[ (\Gamma-i \Omega )^3 \Sigma  \right] }{  \Omega \Gamma (\Gamma^2 + \Omega^2) |\Sigma|^2} .
\end{aligned}
\label{Eq-Pres}
\ee
Here
\begin{equation}
    \Sigma = s_1 \cot{q_1 L_1/2}+s_2 \cot{q_2 L_2/2},
    \label{sigma}
\end{equation}
depends on  complex wave vectors,
\be
q_{1,2} = \frac{\sqrt{\omega(\omega+ i \gamma)}}{s_{1,2}} = \frac{\Omega+ i \Gamma}{s_{1,2}},
\label{Eq-wv}
\ee
where    $\Omega $ and $  \Gamma$ describe, respectively, frequency and damping  of density oscillations. 
 The  complex wave vectors   are different in regions ``1'' and ``2'' ($q_1 \neq q_2$)  due to different plasma velocities $s_{1,2}$, while    $\Omega$ and $\Gamma$ are the same in both regions if $\gamma$ is constant across the PC.

In the resonant regime, when  $\gamma$ is small, the    dissipation $P$  shows sharp   maxima, the positions of which can be found by sending  $\gamma \to  0.$  Then   $\Sigma$  becomes real  and has exact zeros at frequencies that determine plasmonic resonances.

 We also note that 
\be
\begin{aligned}
&P (s_2 = s_1 ) = \frac{ \gamma^2 }{\omega^2 + \gamma^2}  P_0(s_2 = s_1) \\
& = 
P_{\rm Dr} = \frac{ \gamma F_0^2 C s_1^2}{2 e^2 \left(\omega^2 + \gamma^2\right)}.
\end{aligned}
\label{Eq-Drude}
\ee



\subsection{ 
Damping of plasma waves
} 
\label{sec-wave_vec} 
 Wave vectors $q_{1,2}$ depend on plasma wave velocities $s_{1,2}$ 
 and have real and imaginaty parts determined by 
$\Omega = \Omega(\omega, \gamma)$ and $\Gamma = \Gamma(\omega, \gamma),$ respectively.
The imaginary parts  determine  spatial rate of plasma oscillations damping so we introduce the characteristic lengths 
$L_{1,2}^* = {s_{1,2}}/{\Gamma}.$
%
From Eq.~\eqref{Eq-wv}, we find 
 \be
\begin{aligned}
& \Omega = 
\left[ \frac{\gamma^2/2}{\sqrt{1+\gamma^2/\omega^2}-1} \right]^{1/2} \!\!\!\!\approx \begin{cases}
\omega &~\text{for}~ \omega \gg \gamma, \\
\sqrt{\omega \gamma/2} & ~\text{for}~\omega \ll \gamma,
    \end{cases}
\end{aligned}
\label{Eq-Omega}
\ee
\be
\begin{aligned}
&\Gamma 
= \left[ \frac{\gamma^2/2}{\sqrt{1+\gamma^2/\omega^2}+1} \right]^{1/2} \approx \begin{cases}
\gamma/2 & ~\text{for}~\omega \gg \gamma, \\
\sqrt{\omega \gamma/2}  &~\text{for}~ \omega \ll \gamma.
    \end{cases}
\end{aligned}
\label{Eq-Gamma}
\ee
Hence, 
$q_{1,2} \approx \left( 1+ i \right){\sqrt{\omega \gamma/2}}/{s_{1,2}} $ 
for   low frequency, $\omega \ll \gamma,$  and 
$q_{1,2} \approx ({\omega+ i \gamma/2})/{s_{1,2}},$ for high frequency, $\omega \gg \gamma.$  Therefore,  the dissipation lengths read
\be
\begin{aligned}
L_{1,2}^* 
\approx \begin{cases}
2 s_{1,2}/\gamma & ~\text{for}~\omega \gg \gamma, \\
\sqrt{2} s_{1.2}/\sqrt{\omega \gamma}  & ~\text{for}~\omega \ll \gamma.
    \end{cases}
\end{aligned}
\label{L12*}
\ee
Both  $1$ and $2$  regions  can be ``{\it long}'' or ``{\it short}'' depending on the  relation between  $L_i$ and $L_i^*,$ $i=(1,2).$
For {\it long region}  ($L_i^* \ll L_{i}$, i.e. $\omega \gg  \omega_{i}^2/\gamma$) plasma waves rapidly decay from boundaries inside the region $i$.  
For {\it short region} ($L_{i}^* \gg L_{i}$, i.e. $\omega \ll  \omega_{i}^2/\gamma$) the plasma wave also decays exponentially, but the length of decay is much larger than length of strip, so amplitude of plasma wave doesn't change much. This regime also includes quasi-static regime of operation, when  the external frequency is very low $\omega \ll \omega_2^2/\gamma  < \omega_1^2/\gamma \ll \gamma$, so that both $1$ and $2$ regions turn out to be short. 

Importantly, for   $s_1 \gg s_2$  and  $L_1 \sim L_2$ there is  an {\it intermediate } regime, when $L_1^* \gg L_1 $ and at the same time $L_2^* \ll L_2$ ($ \omega_2^2/\gamma \ll \omega \ll \omega_1^2/\gamma$). For such case,  plasmonic oscillations in regions $1$ have approximately constant amplitude but rapidly decay into regions $2,$ so that  plasmonic oscillations in different regions $1$ are effectively disconnected (hence, terms ``active'' and ``passive'' with respect to stripes $1$ and $2$).   In this regime, there are two competing  contributions to the  dissipation, from plasmonic oscillations in regions $1$ and also from narrow  boundary regions $\sim L_2^*$ in the stripes $2.$  As we  will see, in some excitation regimes the second contribution can dominate.

\begin{figure}[h!]
\centering
\includegraphics[width=8.6 cm]{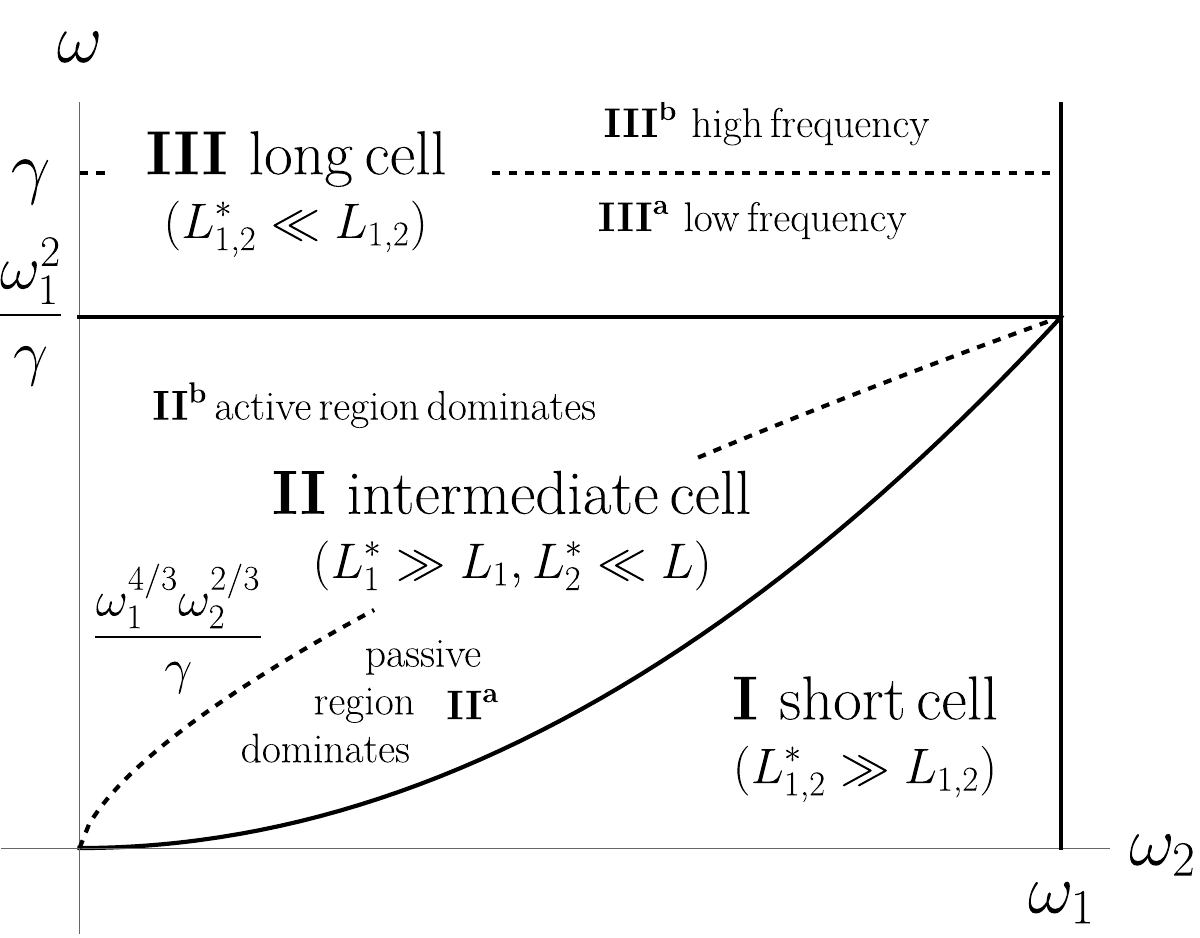}
\caption{Different regimes of dissipation in nonresonant PC ($\gamma \gg \omega_{1,2}$) schematically presented on a $(\omega,\omega_2)$-plane while $\gamma$ and $\omega_1$ are fixed and $\gamma \gg \omega_1 \ge \omega_2$. Here we assume $L_1 \sim L_2$ and omit numerical coefficients. Thick lines divide diagram into three regimes: (I) short cell, (II) intermediate cell and (III) long cell. Condition $L_2^* = L_2$ defines   boundary between (I) and (II) ($\omega = \omega_2^2/\gamma$).  Analogically, line between (II) and (III) is  found from $L_1^* = L_1$ ($\omega = \omega_1^2/\gamma$).
Dashed line $\omega  = \gamma$  splits regime (III) into high and low frequency regimes.  Dashed line $\omega^3 = \omega_1^4 \omega_2^3/\gamma^3$   separates region II into  sub-regions of  (II$^b$)  and  (II$^a$),  where   dominant contribution to the dissipation is given by  plasmonic oscillations  in the active region   or  exponentially decaying tails in the  passive  regions, respectively.}
\label{Fig-nonres_regime}
\end{figure}

\section{Frequency dependence of dissipation}
Next, we discuss evolution of the dissipation with increasing of excitation frequency. 

As follows from Eq.~\eqref{L12*}, lengths $L_{1,2}^*$ decay with frequency. At very low frequency,   both  1 and 2  regions are short, $L_1^* \gg L_1  ,$  $L_2^* \gg L_2. $  We will call this regime as ``short cell'' regime.  In this quasi-static regime, the  system dissipates according to
the static  Ohm’s law.  In the opposite limit of very high frequency  $L_1^* \ll  L_1 , $  $L_2^* \ll L_2, $  plasma oscillations are overdamped and are  spatially limited  to very narrow layers on the inter-regions boundaries. In this case, which we will call ``long sell'' regime,    the main contribution to the dissipation is related to homogeneous Drude dissipation in the external field.   
Below, we discuss  in more detail these  regimes as well as a number of other intermediate regimes.

Linearization of Eqs.~ \eqref{Eq-Navier_Stokes} and \eqref{Eq-continuity} leads to the following  equation, that gives position of plasmonic resonances in the response to  an external force of frequency $\omega$:
\begin{equation}
    \left[\omega^2-\omega_q^2 \right]^2 \tau^2 + \omega^2= 0.
\end{equation}
Here $\omega_q = q s$ and $\tau = \gamma^{-1}$. For a stripe of width $L,$  wave vector $q$ is quantized and the fundamental frequency is determined  by $q=\pi/L.$  For large $\tau,$ this equation gives $\omega \approx \omega_q -i\gamma/2.$
However,   in the non-resonant regime, when $\omega \tau \ll 1$ and $\omega_q \tau \ll 1,$ this equation determines the  Maxwell   relaxation rate instead of frequency of plasmonic resonances:   
\begin{equation}
    \omega = \pm i \omega_q^2 \tau.
\end{equation}
Hence, two relaxation  rates appear in the problem 
\begin{equation}
    \Gamma_{1,2} = \omega_{q_{1,2}
}^2 /\gamma \ll \gamma.
\end{equation}
These rates  have a physical meaning of the Maxwell relaxation rates in the passive and active regions.

\subsection{``Short cell'' regime (see region $\textbf{\rm I}$ at Fig.\ref{Fig-nonres_regime} and Fig.\ref{Fig-5regimes})}
In the quasi-static regime, we have 
\begin{equation}
    \Omega \approx \Gamma \approx \sqrt{\omega \gamma/2}, \, \, \,  \Sigma \approx \frac{\sqrt{2}(1-i)}{\sqrt{\omega \gamma}} \left( \frac{s_1^2}{L_1}+\frac{s_2^2}{L_2} \right) \!, 
\end{equation}
and $L_1^* \gg L_1  ,$  $L_2^* \gg L_2. $
Using these formulas, one   can expand Eq.~\eqref{Eq-Pres} over $\omega$ up to the second order.  The results reads (for simplicity here we take $L_1 = L_2 = L/2$):
\begin{equation}
    P = P_0 \left[\frac{2 s_1 s_2}{s_1^2+s_2^2}\right]^2  \left[1 + \omega^2  \xi \frac{\gamma^2 (1-k^2)^2}{\omega_2^4} -\frac{\omega^2}{\gamma^2}\right].
    \label{Eq-P1}
\end{equation}
Here $k = s_2/s_1$ ($0 \leq k \leq 1$) and  $\xi = \pi^4 (k^4+22 k^2 +1) /2880 (1+k^2)^2$ ($0.03 < \xi < 0.21$).  The first term in the square brackets corresponds to the quasi-static Ohn's law, while term   $\propto \gamma^{-2}$ comes from the  expansion of the  Drude factor:  ${\gamma^2}/({\gamma^2+\omega^2}) \approx 1-{\omega^2}/{\gamma^2}$. 
It worth noting that dissipation at zero frequency strongly depends on type of LPC coupling:  for weak coupling ($s_1\approx s_2$) $P(0) \approx P_0,$ while  for strong coupling  ($s_1\gg s_2$), dissipation is much lower due to blocking of dc current in the zero-frequency limit by low-conducting passive regions:  $P(0) \approx P_0 s_2^2/s_1^2$ .

The term $\propto \xi $  describes retardation  related to the Maxwell relaxation in the passive stripe.  We see that there are two corrections, which are  proprtional to $\omega^2$ and have different signs. For $s_1 \sim s_2$ ($k \sim 1$) the  term   $\propto \xi $ is larger and the  $\omega^2-$correction is positive.  However, in  the weak coupling regime, when $k$ turns to unity, the  $\omega^2-$correction can change sign.          
Assuming now  arbitrary $L_1$ and $L_2$ and denoting   $\delta s = s_1 - s_2$, for weak coupling regime $\delta s \ll s_1,$ the relative correction  is
$$\frac{P(\omega)-P(0)}{P(0)} =  -\frac{\omega^2}{\gamma^2} +\frac{\delta s^2 \gamma^2 L_1^2 L_2^2 (L_1^2+4 L_1 L_2 + L_2^2)}{180 (L_1+L_2)^2 s_1^6 }\omega^2.$$
For $L_1 = L_2 = L/2$ change of sign happens for plasma wave velocity modulation $\delta s/s_1 =2 \sqrt{30} \omega_1^2/\gamma^2 \pi^3 \approx 1.1 ~\omega_1^2/\gamma^2$.

\subsection{``Intermediate cell'' regime 
(see region $\textbf{\rm II}$ at Fig.\ref{Fig-nonres_regime} and Fig.\ref{Fig-5regimes})}
In this regime
\begin{equation}
    \Omega \approx \Gamma \approx \sqrt{\omega \gamma/2}, \, \, \,  \Sigma \approx -i s_2 + \frac{(1-i)s_1^2}{\sqrt{\omega \gamma} L_1}, 
\end{equation}
$L_2^* \ll L_2,$  $L_1^* \gg L_1,$ 
and plasma waves existing in the active regions  exponentially decay into region ``2'' from the inter-region  boundaries. Conditions on frequency of intermediate cell regime read  $\omega_2 \ll \sqrt{\omega \gamma} \ll \omega_1.$   Keeping in mind that $\omega \ll \gamma$ we find from  Eq.~\eqref{Eq-Pres}
\be
\begin{aligned}
&P =P_0  \frac{\pi^4}{120}\frac{ \omega^2 \gamma^2 L_1}{ \omega_1^4 (L_1+L_2)}
\\
&+P_0 \left[\frac{(2 L_1 + L_2) s_2^2}{L_1 s_1^2 + L_2 s_2^2} + \frac{\sqrt{\omega \gamma} L_1^2 s_2 }{2 \sqrt{2} (L_1 s_1^2 + L_2 s_2^2) }\right]
\end{aligned}
\label{Eq-inter_cell}
\ee
First term, proportional to  $ \omega^2,$  represents contribution  to the dissipation from the  active region, where $L_1^* \gg L_1$.  Physically, this term is responsible for retardation effect  related   to the Maxwell relaxation in the active regions.  
The  frequency-dependent term,  $\propto \sqrt{\omega},$ in the square brackets   describes dissipation within boundary regions on the order $L_2^*$     in  the passive regions.    These two terms compete with each other, so that  
either passive or active region can    dominate in the dissipation.
Equalizing the terms proprotional to $\omega^2$ and $ \sqrt{\omega},$ and assuming $L_1 = L_2 =L/2$ 
one gets   $$\omega  =\omega^*= \left(\frac{1800 \, \omega_1^4 \omega_2^2}{\pi^6 \gamma^3} \right )^{1/3} 
\approx 1.24 \, \frac{\omega_1^{4/3} \omega_2^{2/3}}{\gamma} . $$
 For $\omega \gg \omega^*$  the term,    proportional to $ \omega^2,$ is larger, so that dissipation in active region  gives the leading contribution   (see region $\textbf{\rm II}^\textbf{b}$ at Fig.\ref{Fig-nonres_regime} and Fig.\ref{Fig-5regimes}). The contribution of   passive regions can be neglected and the system without loss of generality can be considered  as an array of active strips separated by insulating regions.
 
 For smaller frequency $\omega \ll \omega^*,$ 
 the  term which is $ \propto \sqrt \omega$ dominates.
 This contribution comes from exponentially decaying plasma waves in region ''2''.

\subsection{ ``Long cell'' regime, (see region $\textbf{\rm III}$ at Fig.\ref{Fig-nonres_regime} and Fig.\ref{Fig-5regimes})}

In this case, plasma waves exponentially decay in both active and passive regions: $L_{1,2}^* \ll L_{1,2}$, i.e. $\omega \gg \omega_{1,2}^2/\gamma$. 
Physically this means, that in both regions the time of the Maxwell  relaxation  becomes larger than the period of the external field oscillation.   
The dissipation  in this case is given by 
the following equation
\be
\begin{aligned}
&P = \frac{ P_0 \gamma^2}{\omega^2+\gamma^2 } \left[ 1 - \frac{(s_1-s_2)^2 (s_1+s_2) (3 \Gamma^2 - \Omega^2)}{\Gamma (\Gamma^2+\Omega^2)(L_1 s_1^2+L_2 s_2^2)} \right],
\end{aligned}
\label{Eq-P_long}
\ee
which is the product of the Drude dissipation  in the homogeneous  external field and a factor in square brackets, which   slowly increase with $\omega$ due to correction caused by finite length of the boundary layer. The  latter is negative at relatively low frequency  and saturates  at positive value for very high frequencies, where $\omega$ becomes larger than  $\gamma,$ so that  plasmon damping length saturates on the value $s_1/\gamma$ for $\omega \to \infty.$     
 Hence,  there are two regions  $\textbf{\rm III}^\textbf{a}$ and $\textbf{\rm III}^\textbf{b}$    at  Figs.\ref{Fig-nonres_regime} and \ref{Fig-5regimes}.

\subsubsection{Low frequency, region $\textbf{\rm III}^\textbf{a}$ at Fig.\ref{Fig-nonres_regime} and Fig.\ref{Fig-5regimes}}
 For  $ \omega_1^2/\gamma  \ll \omega \ll \gamma,$  we get 
\begin{equation}
    \Omega \approx \Gamma \approx \sqrt{\omega \gamma/2}, \, \, \,  \Sigma \approx -i (s_2 + s_2)
\end{equation}
and  Eq.\ref{Eq-P_long} simplifies:
\begin{equation}
    P \approx P_0 \left[ 1- \frac{ \sqrt{2} (s_1 - s_2)^2 (s_1 + s_2)}{ \sqrt{\omega \gamma} (L_1 s_1^2 + L_2 s_2^2) }  -\frac{\omega^2}{\gamma^2}\right].
    \label{Eq-P3}
\end{equation}
Two frequency-dependent  terms in the square brackets represent  small corrections,  which yield the maximum of $P$  (see Fig.~\ref{Fig-5regimes}).   For $s_2 \ll s_1,$ dissipation is maximal at $\omega \sim \omega_1^{2/5} \gamma^{3/5}.$    

Introducing 
 $L_{\rm eff} = (L_1 s_1^2+L_2 s_2^2)/s_1^2$ ($L_{\rm eff}|_{s_2 = s_1} = L_1+L_2$, $L_{\rm eff}|_{s_2 = 0} = L_1$), one  can rewrite Eq.~\eqref{Eq-P3} as follows
\begin{equation}
    P = P_0 \left[ 1- \left( 1 - \frac{s_2}{s_1} \right)^2 \frac{L_1^*+L_2^*}{L_{\rm eff}} -\frac{\omega^2}{\gamma^2}\right].
    \label{Eq-P_over_Leff}
\end{equation}
We also note that in the isolated 
 strip ($s_2 = 0$) $P = P_0 (1-L_1^*/L_1  -\omega^2/\gamma^2)$ and in the  homogeneous system ($s_2  =s_1$),  dissipation is given by the Drude formula.
\subsubsection{High frequency, region $\textbf{\rm III}^\textbf{b}$ at Fig.\ref{Fig-nonres_regime} and Fig.\ref{Fig-5regimes}}
\label{Sec-high_freq}
When external frequency exceeds  $\gamma,$  damping of plasma waves saturates   \begin{equation}
    \Omega \approx \omega,\, \, \, \Gamma \approx \gamma/2, \, \, \,  \Sigma \approx -i (s_2 + s_2).
\end{equation}
and Eq.~\eqref{Eq-P_long} becomes
\begin{equation}
    P = \frac{\gamma^2}{\omega^2} P_0 \left[ 1 + \frac{2  (s_1-s_2)^2 (s_1+s_2)}{\gamma (L_1 s_1^2 + L_2 s_2^2)} \right].
    \label{Eq-P7}
\end{equation} 
It is worth noting that the second term in the square brackets is small frequency-independent correction which have opposite sign to $1/\sqrt \omega \, \,  -$ correction in  Eq.~\eqref{Eq-P3}. This correction comes from the narrow region of width $s/\gamma$ near  the boundary.

\section{Discussion and conclusion}
\begin{figure}[h!]
\centering
\includegraphics[width=8.6 cm]{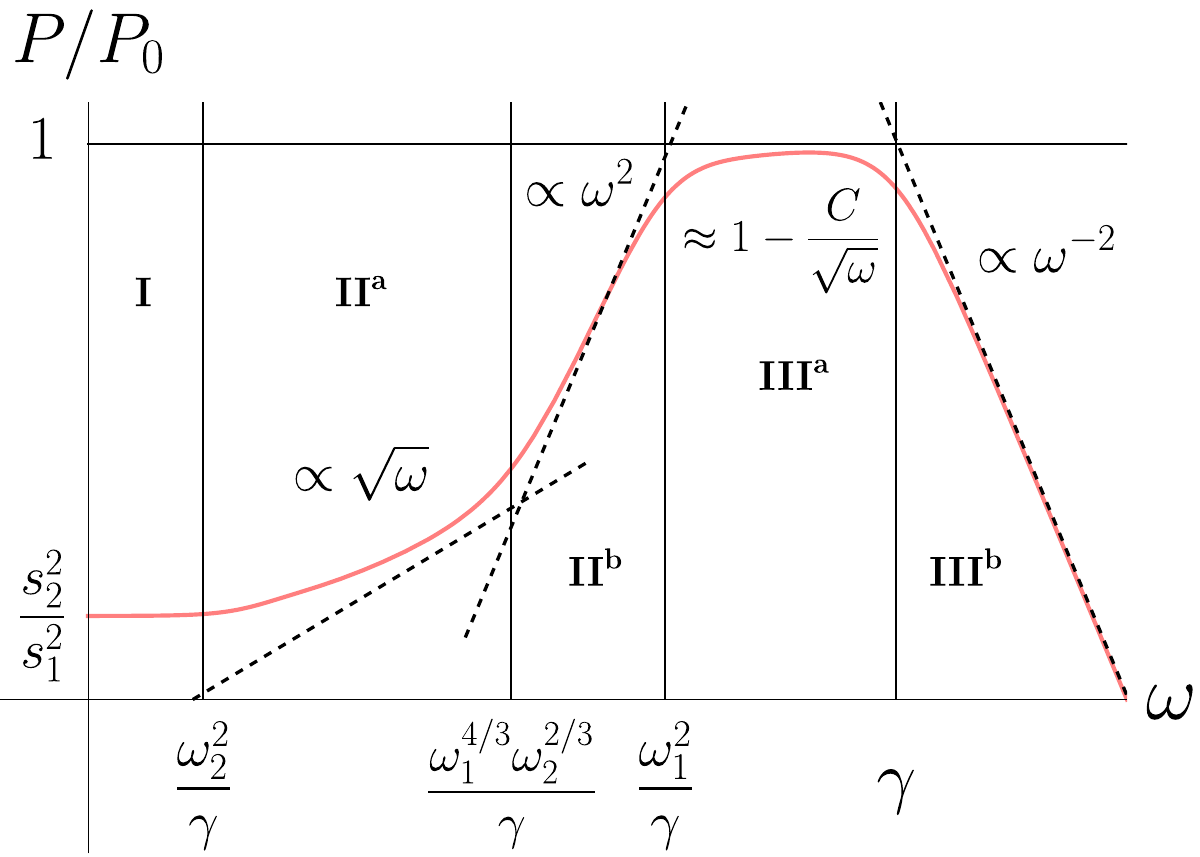}
\caption{
Different regimes of dissipation in nonresonant PC ($\gamma \gg \omega_{1,2}$) at log-log plot of normalized dissipation on frequency (red line). Frequencies labeled at horizontal axis together with vertical dashed lines represent boundaries of described regimes. Value $P/P_0 = s_2^2/s_1^2$ represents approximation of zero-frequency dissipation in strong coupling regime. With dashed lines at regions $\textbf{\rm II}^{\textbf{a}}$, $\textbf{\rm II}^{\textbf{b}}$ and $\textbf{\rm III}^{\textbf{b}}$ we draw asymptotics $\sqrt{\omega}$, $\omega^2$ and $\omega^{-2}$ correspondingly. We omit numerical values when label both axis; dissipation is plotted using $L_1 = L_2, \, \omega_2/\omega_1 = 0.01, \,  \omega_1/\gamma = 0.1$; $C \approx 2 \sqrt{2} s_1/L\sqrt{\gamma}$
}
\label{Fig-5regimes}
\end{figure}

Analyzing results obtained above we conclude that the most interesting effects are related to the Maxwell relaxation. Remarkably, in the strongly dissipative regime under discussion, the rates of the Maxwell relaxation    are much smaller than the dissipation rate: $\Gamma_{1,2} \ll \gamma.$ On the other hand, these rates describe characteristic scales of variation $P$ with frequency. In particular,  in the region II$^b,$ dissipation sharply increase with $\omega$ within the frequency interval  on the order of $\Gamma_1 \ll  \gamma.$  Consequently,  $P(\omega)$ has a sharp minimum with a width $\sim \Gamma_1$, and this resonance is better seen if we re-plot Fig.~\ref{Fig-5regimes} in normal (not log-log) scale.    

To conclude, we have studied  the  transmission  of the THz radiation through the  lateral plasmonic
superlattice with a unit cell consisting from two regions with different plasma wave velocities. We have developed a theory of the  non-resonant absorption and described evolution  with increasing radiation frequency. Several absorption  regimes  are identified, described analytically and illustrated on the general diagram.  It is demonstrated that although being deeply  in the non-resonant regime, the system in strong coupling shows very sharp dependence on the frequency within the very small frequency scale, $\omega_1^2/\gamma$  determined by the  Maxwell relaxation, which is  much smaller than the damping rate $\gamma.$ The responsivity of the system with respect to gate voltages strongly increases in this frequency domain.

\newpage

\section*{Acknowledgements}

We thank S. L. Rumyantsev for very useful discussions. 
The work  was supported by the 
Russian Science Foundation under grant 24-62-00010.
The work of I.G. was also partially supported by the Theoretical Physics and Mathematics Advancement Foundation ``BASIS''.

\bibliography{main}

\end{document}